\documentclass[a4paper,11pt]{article}
\usepackage{graphicx}
\usepackage{epsfig}
  
\newcommand{\ds}{\displaystyle }

\newcommand{\N}{{\sf N\hspace*{-0.9ex}%
\rule{0.15ex}{1.5ex}\hspace*{0.9ex}}}

\title{Fractional random walk lattice dynamics}

\author{ {\sl T.M. Michelitsch$^{1}$\footnote{Corresponding author, e-mail~: michel@lmm.jussieu.fr },
B.A. Collet$^{1}$}, A.P. Riascos$^2$ \\ \\ {\sl A. F. Nowakowski$^3$, F.C.G.A. Nicolleau$^3$}
\\ \\
$^1$ Sorbonne Universit\'es \\ Universit\'e Pierre et Marie Curie (Paris 6) \\ Institut Jean le Rond d'Alembert, CNRS UMR 7190 \\
4 place Jussieu, 75252 Paris cedex 05, France
\\ \\
$^2$Instituto de F\'{i}sica, Universidad Nacional Aut\'{o}noma de M\'{e}xico\\  Apartado Postal 20-364, 01000 M\'{e}xico, D.F., M\'{e}xico
\\ \\
$^3$ Sheffield Fluid Mechanics Group\\
Department of Mechanical Engineering\\
University of Sheffield\\
Mappin Street, Sheffield S1 3JD,
United Kingdom
\\ \\ {\it Submitted manuscript} %\\ \\ {\small \it \jobname .tex }
}
\begin{document}

\maketitle

\newpage

\begin{abstract}
% BELOW SHOULD BE THE TEXT OF YOUR ABSTRACT
We analyze time-discrete and continuous `fractional' random walks on undirected regular networks with special focus on cubic periodic lattices in $n=1,2,3,..$ dimensions.
The fractional random walk dynamics is governed by a master equation involving {\it fractional powers of Laplacian matrices $L^{\frac{\alpha}{2}}$}
where $\alpha=2$ recovers the normal walk.
First we demonstrate that
the interval $0<\alpha\leq 2$ is admissible for the fractional random walk.
 We derive analytical expressions for fractional transition matrix and closely related the average return probabilities. We further obtain the
fundamental matrix $Z^{(\alpha)}$, and the mean relaxation time (Kemeny constant) for the fractional random walk.
The representation for the fundamental matrix $Z^{(\alpha)}$ relates fractional random walks with normal random walks.
We show that the fractional transition matrix elements exihibit for large cubic $n$-dimensional lattices a power law decay of an $n$-dimensional infinite space
Riesz fractional derivative type indicating
emergence of L\'evy flights. As a further footprint of L\'evy flights in the $n$-dimensional space, the fractional transition matrix and fractional return probabilities
are dominated for large times $t$ by slowly relaxing long-wave modes
leading to a characteristic $t^{-\frac{n}{\alpha}}$-decay.
It can be concluded that, due to long range moves of fractional random walk, a small world property is emerging increasing the efficiency to explore the lattice
when instead of a normal
random walk a fractional random walk is chosen.
\end{abstract}

\section{Introduction}

The study of dynamical processes on networks has become a huge interdisciplinary research area during the last decade \cite{newmann}.
The characteristic features of many
natural systems such as biological, social and computer systems (the world wide web) \cite{zhang}, and last but not least the molecular structure of materials can be described as complex networks \cite{NohRieger}.
There is a huge amount of works
addressing the random motions on networks and many models have been developed
to capture the structural features of random motions on networks \cite{dorogotsev,Watts,zhang}.
In order to analyze dynamical processes on networks,
it has been demonstrated that random walks are a highly useful concept to describe problems such as of exploration, search, navigation and propagation of information on
networks \cite{beni}. Whereas normal random walks with emerging Brownian motions have been widely used to describe the dynamics on networks, e.g. \cite{NohRieger,zhang,doyle},
and many others,
it has turned out that many
processes associated with long-range jumps on the network, cannot be properly described by normal random walks allowing in a time-step only moves to connected nodes.
Many complex phenomena such as anomalous transport phenomena with long-range jumps in diffusion processes indeed
cannot be described by Brownian motions. However, their power-law long-range jump characteristics can be described by L\'evy motions \cite{metzler-checkin,checkin}
in the framework of L\'evy statistics where such stochastic
processes include long-range moves and are in many cases described by continuous space fractional operators \cite{metzler,metzler2014}.
Further diffusion processes have been analyzed governed by a time-fractional Fokker Planck equation
\cite{Barkai2001}.

The fractional calculus in the continuous space has become a standard tool to describe a large ensemble of interdisciplinary problems of complex behavior
involving effects due to spatial long-range interactions. A prominent example is for instance the development of fractional quantum mechanics by Laskin \cite{Laskin}.
Depending on the functional spaces where these fractional operators are defined we find
various definitions of fractional operators in the continuous space in the literature  \cite{hilfer-2008,metzler,samko,samko2003,podlubny}.
Whereas these fractional operators
are obtained as convolutional integrals in a continuous space, the fractional calculus on discrete networks and lattices is less developed.
An approach to define fractional differential operators on lattices is suggested by Tarasov \cite{tara1}.
Beside the applications on diffusion problems on the lattice, the importance of fractional lattice models appears also for a
description of fractional lattice vibrational phenomena, a generalization of crystal lattice dynamics. Some initial steps towards such a fractional
generalization generalization of nonlinear classical lattice dynamics has been introduced by
Laskin and Zaslavsky \cite{laskin2006}. In a lattice dynamics model which defines by Hamilton's variational principle the `Laplacian matrix'
which contains all constitutive
information of the harmonic interparticle interactions, it is therefore desirable to develop a
`fractional generalization' of the existing lattice dynamics approach \cite{michel-riascos-et-al-APM,michel-collet}.

In the context of Markovian processes on networks, the concept of `fractional random walk on undirected networks' generalizing the `normal random walk'
was recently introduced
by Riascos and Mateos
\cite{riascos12,riascos-fracdyn,riascos-fracdiff,riascos15}. In contrast to the normal random walk where the walker can reach in one time-step only connected nodes,
the fractional random walker is
allowed to reach in one time-step any node in a connected network introducing a small world property to the network \cite{riascos-fracdyn,riascos-fracdiff}.
It was found for different type of complex networks numerically, and for the cyclic ring explicitly,
that the efficiency to navigate on networks is increased when the fractional random walk is chosen
instead of a normal walk.
Moreover, the power law decay of the fractional Laplacian matrix on sufficiently large 1D cyclic rings (periodic one-dimensional lattices)
indicates in the context of fractional diffusion the emergence of L\'evy flights,
 \cite{riascos-fracdyn,riascos-fracdiff} and see also the analysis in \cite{michel-riascos-et-al-APM}.

Inspired by the above findings where the fractional generalization of a network Laplacian matrix introduces a small world property to large world networks, the fractional
random walk dynamics opens a huge new interdisciplinary field which merits further studies. It is now highly desirable to better understand the mechanism of
`fast fractional navigation' in a network. There is a huge potential of applications in interdisciplinary
research areas as various as search strategies on computer networks such as the world wide web, the propagation of information
in societies, the spread of pandemic diseases, of cancer cells in the body, etc.
\\ \\
The goal of the present paper is to derive a theoretical framework
to describe fractional random walk dynamics on undirected regular networks with main focus on $n$-dimensional cubic periodic and infinite lattices.
The advantage of these regular network systems is that they accessible to an explicit spectral analysis as the eigenspace is terms of Bloch-vectors is explicitly available.
In the present paper the fractional random walk versus normal random walk is analyzed by means of a fractional generalization of master equation.
 For the sake of simplicity and analytical accessibility, we confine us in the present paper
on rather simple species of regular undirected networks with constant degree for all nodes with special focus on
$n$-dimensional periodic cubic lattices. In the $n$-dimensional cubic lattice we assume only next neighbor node connections, i.e. in any physical
dimension $j=1,..,n$ each node has two connected neighbor nodes. Thus the degree
of all nodes is identically equal to $2n$. Further we assume that the cubic lattice is periodic in any spatial dimension $j=1,..,n$
where the spatial dimension $n=1,2,3,4,..\in \N$.
Despite of these restrictions, many derivations of this paper can be easily generalized to more complex undirected networks.

The present paper is organized as follows.
In subsequent section \ref{sec2} we define a `good', physically admissible fractional Laplacian matrix for undirected regular and $n$-dimensional cubic networks.
The fractional Laplacian matrix is generated by a power law matrix function of a `simple' Laplacian matrix involving only next neighbor node connections.
In section \ref{sec3} we evoke briefly some general basic features of time-discrete and time-continuous random walks on regular undirected networks in the framework of Markovian process which we will need for the analysis
of fractional random walks.

To define the `fractional random walk' a `fractional master equation' involving a
fractional Laplacian matrix $L^{\frac{\alpha}{2}}$ as generator matrix for the time evolution is introduced (section 4).
The fractional random walk contains for $\alpha=2$ the normal random walk.
We show explicitly (see appendix) that fractional random walk approach is admissible in the interval $0<\alpha \leq 2$.
We derive expressions for the transition matrix and average return probabilities.
We show that the fractional transition matrix elements exhibit for large cubic $n$-dimensional lattices a power law decay of an $n$-dimensional infinite space
Riesz fractional derivative type indicating
emergence of L\'evy flights. As a further footprint of L\'evy flights in the $n$-dimensional space, the fractional transition matrix and fractional return probabilities
are dominated for large times $t$ by slowly relaxing long-wave modes modes
leading to a characteristic $t^{-\frac{n}{\alpha}}$-decay (sections 4.1 and 4.2).
The asymptotic expressions further recover known results for normal
and fractional random walk on the 1D cyclic ring \cite{riascos-fracdyn,riascos-fracdiff}.

In order to obtain further characteristic information on the fractional random walk we introduce a Green's matrix in section 4.3 which yields a convenient representation of the
fundamental matrix $Z^{(\alpha)}$, containing the fractional mean relaxation time (Kemeny constant) for the fractional random walk.
The representation for the fundamental matrix $Z^{(\alpha)}$ relates fractional random walks with normal random walks (section 4.3).
The analysis of the mean relaxation time (Kemeny constant) demonstrates for the $n$-dimensional cubic lattice the increased efficiency of a fractional random walk strategy.
This analytical result supports the numerical findings of recent works \cite{riascos-fracdyn,riascos-fracdiff}.

\section{Fractional Laplacian matrix on regular undirected networks and cubic lattices}
\label{sec2}

For general derivations performed in this paper,
we consider undirected regular networks of $N$ nodes. Let us denote with $p=0,..,N-1$ the nodes of this network. We assume for simplicity that all nodes of the network have
constant degree $K_p=K$. In order to obtain explicit expressions we specify the network to cubic lattices with only next neighbor node connections.
We consider the cubic lattice in $n=1,2,3,..,$ dimensions with periodic boundary conditions in any dimension $j=1,..,n$. Topologically such a lattice can be conceived
as $n$-dimensional hypertorus (`$n$-torus'), for instance in 1D this is a cyclic ring, in 2D a conventional torus, and so forth.
We assume the lattice contains $N=N_1\times ..\times N_n$ lattice points where each lattice point can be identified with a node of the network.
The lattice is assumed to be $N_j$-periodic in any spatial dimension $j=1,..,n$.
For the cubic lattice the nodes are denoted by the vector
$\vec{p}=(p_1,p_2,..,p_n)$ ($p_j=0,..N_j-1$). $N\times N$-matrices ${\cal M}$ defined on the cubic lattice are denoted by
${\cal M}_{\vec{p}\vec{q}} =: {\cal M}_{p_1,..,p_n|q_1,..,q_n} $.
In order to define the lattice
(fractional) Laplacian matrices, it is sufficient to confine ourselves to scalar fields $u_{\vec{p}}$ (one field degree of freedom) associated to each node
of first of all unspecified physical nature.

Before we define fractional Laplacian matrices, let us first define a `good' i.e. physically admissible Laplacian matrix on a undirected network.
To this end we introduce a potential energy in the form of a positive (semi-) definite quadratic form

\begin{equation}
 \label{pos}
 V= \frac{1}{2}\sum_{p=0}^{N-1}\sum_{q=0}^{N-1} u_q^* L_{pq} u_p \geq 0
\end{equation}
where on an undirected network the $N\times N$-Laplacian matrix $L_{pq}=L_{qp}$ is symmetric. Assuming identical degree $K=K_p$ for all nodes, it
can be represented as
\begin{equation}
 \label{repres}
 L_{pq} =K \delta_{pq}-A_{pq}
\end{equation}
where $A_{pq}$ denotes the adjacency matrix with $A_{pq}=1$ ($A_{pq}=0$) if node $p$ is (not) connected with node $q$ and the degree  $K_p=K=\sum_{q=0}^{N-1}A_{pq}$
counts the number of connections of a node $p$. We assume further that $A_{pp}=0 \,\,\forall p=0,..,N-1$.
In the approach to be developed it is sufficient to consider a `large world network'  Laplacian matrix which we define as $\frac{K}{N} << 1$,
i.e. the degree (number of connections of a node) is much less than the total number of $N$
nodes. It follows that translational symmetry $\sum_{q=0}^{N-1} L_{pq} = 0$ is fulfilled,
which means that the constant vector with the components
$u_p=1$ is eigenvector of the Laplacian matrix with the only vanishing eigenvalue $\lambda_0=0$ [which we denote
throughout this paper by $\lambda_0$]. Further, the requirement of elastic stability of (\ref{pos}) requires that $N-1$ eigenvalues of the Laplacian matrix
$\lambda_{\ell} >0$ $\ell=1,..N-1$ are positive. The sign convention of the Laplacian matrix is chosen here such that the Laplacian matrix is
the matrix-analogue to $-\frac{d^2}{dx^2}$, namely as a positive semi-definite operator.
For a discussion we refer to \cite{michelJphysA,michel-collet}
\footnote{Contrarily to the convention in these works where the Laplacian matrix
is defined negative semi-definite matrix-analogue to the Laplacian operator $\frac{d^2}{dx^2}$.}
The Laplacian matrix of a cubic lattice with only next neighbor node connections is defined as the form
\begin{equation}
 \label{BvK}
 {\cal L} = \left(2n {\hat 1}-A_n \right)
\end{equation}
where ${\hat 1}$ denotes the $N\times N$ unity matrix. The components of the Laplacian matrix can be represented as
\begin{equation}
\label{compo}
{\cal L}_{\vec{p}\vec{q}} = 2n \delta_{\vec{p}\,\vec{q}} -A_{\vec{p}\,\vec{q}}\, .
\end{equation}
This representation needs to be read by introducing the notations $\delta_{\vec{p}\vec{q}}= \prod_{j=1}^n\delta_{p_jq_j}$ for the components of the unity matrix on the cubic lattice with
$\vec{p}=(p_1,..,p_n)$, $\vec{q}=(q_1,..,q_n)$ where $\delta_{ij}$ denotes the Kronecker symbol.
The adjacency matrix for the cubic lattice with only next neighbor node connections can be represented by
\begin{equation}
\label{cubicadj}
A_{\vec{p}\vec{q}} = \sum_{j=1}^n [D_j+D_j^{\dagger}]_{\vec{p}\vec{q}} ,
\end{equation}
where $D_j$ and $D_j^{\dagger}=D_j^{-1}$ denote the (on the periodic and infinite lattice unitary) next neighbor shift operators in the $j=1,..,n$-directions
defined by $D_ju_{p_1,..p_j,..,p_n}= u_{p_1,..p_j+1,..,p_n}$
and  its adjoint (inverse) shift $D_j^{\dagger}u_{p_1,..p_j,..,p_n}= u_{p_1,..p_j-1,..,p_n}$.
On the periodic and infinite lattices the shift operators are unitary \cite{michel-collet}.
$D_j$ shifts the field associated to lattice point $\vec{p}=(..,p_j,..)$
to the field associated with the adjacent lattice point in the positive $j$-direction $(..,p_{j+1},..)$. The inverse (adjoint) shift operator
$D_j^{\dagger}=D_j^{-1}$ shifts to the adjacent lattice point in the negative $j$-direction $(..,p_{j-1},..)$. For the $n$-dimensional cubic lattice the components of the
shift operators can be represented by

\begin{equation}
 \label{compis}
 [D_j]_{\vec{p}\vec{q}} =  \delta_{p_{j+1}q_j}    \prod_{s\neq j}^n \delta_{p_sq_s} ,\hspace{1cm}
 [D_j^{\dagger}]_{\vec{p}\vec{q}} =  \delta_{p_{j-1}q_j}    \prod_{s\neq j}^n \delta_{p_sq_s}
\end{equation}
and with (\ref{compis}), the components of the Laplacian (\ref{compo}) of the cubic lattice with next neighbor node connections are
\begin{equation}
 \label{explicit}
 {\cal L}_{p_1,..p_n|q_1,..,q_n} =2n \prod_{j=1}^n\delta_{p_jq_j} - \sum_{j=1}^n\left(\delta_{p_{j+1}q_j}+ \delta_{p_{j-1}q_j} \right)\prod_{s\neq j}^n \delta_{p_sq_s} .
\end{equation}
The Laplacian matrix (\ref{repres}) has a spectral representation of the form
\begin{equation}
 \label{spectrallap}
 L = \sum_{\ell=0}^{N-1} \lambda_{\ell}|\ell\rangle\langle\ell|
\end{equation}
where we employed Dirac's bra-ket notation for the representation of the eigenvectors. The eigenvector $|0\rangle$ refers to the vanishing eigenvalue $\lambda_0=0$ and has constant components $[|0\rangle]_p = \frac{1}{\sqrt{N}} \forall p$ reflecting translational symmetry.
Due to the symmetry (self-adjointness) of the Laplacian matrix $L_{pq}=L_{qp}$, we have always the property of ortho-normality of the eigenvectors
\footnote{I. e. $\langle\ell|j \rangle =\delta_{\ell j} $ and ${\hat 1} = \sum_{\ell=0}^{N-1}
|\ell\rangle\langle\ell|$.}.
With these preliminary comments we can define the fractional Laplacian matrix by its spectral representation as a power matrix function

\begin{equation}
\label{fraclapower}
 L^{\frac{\alpha}{2}} = \sum_{\ell=0}^{N-1} \lambda_{\ell}^{\frac{\alpha}{2}} |\ell\rangle\langle\ell| ,\hspace{2cm} \alpha >0
\end{equation}
having the same good properties as the Laplacian (\ref{spectrallap}). So we introduce the `fractional potential' as
\begin{equation}
 \label{pospotfra}
 V_{\alpha} =\frac{1}{2}\sum_{p=0}^{N-1}\sum_{q=0}^{N-1} u_q^* [L^{\frac{\alpha}{2}}]_{pq} u_p \geq 0
\end{equation}
which is as well a positive semi-definite quadratic form.
For the periodic cubic lattice the field $u_{\vec{p}}$ fulfills the periodicity conditions
\begin{equation}
 \label{periodiccon}
 u_{p_1,..p_j+N_j,..p_n} = u_{p_1,..p_j,..p_n} ,\hspace{1cm} j=1,..,n
\end{equation}
which are also fulfilled by the shift operator matrices (\ref{compis}). This allows cyclic index convention $0\leq p_j \leq N-1 \,\, \forall j=1,..,n$ \cite{michelJphysA}. As a consequence the spectral representation of the fractional Laplacian
matrix  ${\cal L}^{\frac{\alpha}{2}}$ on the cubic lattice can be expressed by Bloch vectors as
\begin{equation}
\label{spectralfracb}
[{\cal L}^{\frac{\alpha}{2}}]_{\vec{p}\vec{q}} = [{\cal L}^{\frac{\alpha}{2}}]_{\vec{p}-\vec{q}}
=\frac{1}{N}
\sum_{\vec{\ell}} e^{i\vec{\kappa}_{\vec{\ell}}\cdot(\vec{p}-\vec{q})}\lambda_{\vec{\ell}}^{\frac{\alpha}{2}} ,
\hspace{0.5cm} \lambda_{\vec{\ell}} = \left(2n-2\sum_{j=1}^n\cos{(\kappa_{\ell_j})}\right) ,\hspace{0.5cm}  \alpha >0
\end{equation}
where we have introduced the more compact notation $\sum_{\vec{\ell}} (..)=\sum_{\ell_1=0}^{N_1-1}..\sum_{\ell_n=0}^{N_n-1}(..)$ and where the sum is performed over the entire set of
$N=\prod_{j=1}^n N_j$ Bloch-vectors $\vec{\kappa}_{\ell}=(\kappa_{\ell_1},.., \kappa_{\ell_n})$ having the
components $\kappa_{\ell_j} =\frac{2\pi}{N_j}\ell_j$ (and $\ell_j=0,..,N_j-1$, with $j=1,..,n$). We see again that only the eigenvalue which corresponds to
the zero wave vector with $\vec{\ell}=(0,..,0)$ is vanishing whereas all other $N-1$
eigenvalues $\lambda_{\vec{\ell}}^{\frac{\alpha}{2}}$ are positive.  For $\alpha=2$ (\ref{spectralfracb}) recovers the spectral representation of the Laplacian matrix with next neighbor node connections (\ref{explicit}).

\section{Some general remarks on random walks on regular undirected networks}
\label{sec3}

Let us define the notion of a ´random walk on the network´ in the framework of Marvovian process.
Let $P_q(t)$ be the probability that the random walker occupies node $q=0,..,N-1$ at a time $t$. The normalization condition
$\sum_{q=0}^{N-1} P_q(t) =1$ indicates that the random walker is at time $t$ for sure somewhere on the network. As we deal with probabilities
we have $0\leq P_q(t) \leq 1$ and the normalization condition is fulfilled at any time $0\leq t < \infty$. It is further convenient to introduce the probability vector
$\vec{P}(t)=[P_0(t), ..P_q(t),..P_{N-1}(t)]$ having the occupation probabilities $P_q(t)$ as components.

The random walk is now defined by the time evolution of the occupation probabilities $P_q(t)$ of the nodes $q$. We assume the random walker is allowed to undertake one discrete move
from one node to another connected node during a given time increment $\delta t$ (time-discrete random walk). We consider each move of the random walk to be a Markovian process
which means, when we decompose the time evolution of the occupation probabilities into discrete steps $\delta t$, that the probabilities
at a time $t+\delta t$ depend only on the probabilities at $t$.
Assuming a linear dependence, this time evolution can be expressed by
 a discrete master equation, namely [where we employ alternatively both, index and matrix notations]
\begin{equation}
 \label{timemaster}
 P_p(t+\delta t) = \sum_{q=0}^{N-1} W_{pq}(\delta t)P_q(t) ,\hspace{2cm} \vec{P}(t+\delta t) =  W(\delta t)\cdot \vec{P}(t)
\end{equation}
where the $N\times N$ matrix $W(\delta t)$ is referred to as transition matrix
connecting the probabilities $\vec{P}(t+\delta t)$ with $\vec{P}(t)$. Since we are considering Markovian processes, transition matrix $W(\delta t)$ in
(\ref{timemaster}) depends only on the time increment $\delta t$ but not on (previous times) $t$. The matrix elements $W_{pq}(\delta t)$
indicate the conditional probability
that the random walker that occupies node $q$ at time $t$ has moved to node $p$ when the time has increased by an increment $t+\delta t$.
It follows from the normalization of the $P_q(t)$ that $W_{pq}(\delta t)$ has to fulfill the normalization condition $\sum_{p=0}^{N-1} W_{pq}(\delta t) =1$,
and since we deal with probabilities
all matrix elements are within $0\leq W_{pq} (\delta t) \leq 1$.
Further we see that the transition matrix after $m$ time-steps is obtained by
\begin{equation}
W(t=m\delta t) = W^m(\delta t) ,\hspace{1cm} m\in \N  \hspace{1cm} W(0)= W^{0}(\delta t) = {\hat 1}.
\end{equation}
 The $W(t=m\delta t)= W^m(\delta t)$ denotes the transition matrix connecting the probability vector at time
$t=m\delta t$ (after $m$ time steps)
$\vec{P}(t=m\delta t)=W(t=m\delta t)\cdot \vec{P}(0)$ with an initial probability vector $\vec{P}(0)$.
We emphasize that normalization and positiveness of the matrix elements of the transition matrix $W(t=m\delta t)$ $\forall m\in \N$
remain conserved for all times $t= m\delta t$
\begin{equation}
\label{norma}
 \sum_{p=0}^{N-1} W_{pq}(t) =1 ,\hspace{2cm} 0 \leq W_{pq}(t) \leq 1\, .
\end{equation}
Note that the normalization condition is equivalent to the fact that the stationary (equal-)
distribution $P_p= \frac{1}{N} \forall p$ is an eigenvector referring to the (largest) eigenvalue $1$ of transition matrix $W(t)$. It is further clear that in order to maintain (\ref{norma})
for all times for $m\rightarrow \infty$ time increments, that all other $N-1$ eigenvalues of the transition matrix must be positive and inferior to $1$ with exponentially
evanescent contributions for increasing number $m$ of time steps.
It follows that $\lim_{t\rightarrow \infty} P_q(t) = \frac{1}{N}$ with
$\lim_{t\rightarrow \infty} W_{pq}(t) = W_{pq}(\infty) = \frac{1}{N}$ $\forall p,q=0,..,N-1$, i.e. all
matrix elements of the stationary transition matrix take the value $\frac{1}{N}$ reflecting the Markov chain convergence theorem \cite{norris}.
Let is now specify the transition matrix $W(\delta t)$ and link it with the properties of the network. We can do so by assuming $W(\delta t)$ to be
generated for a small (in the limiting case infinitesimal) time-step $\delta t$ as
\begin{equation}
 \label{transmat}
 W_{pq}(\delta t) =\delta_{pq} - \delta t L_{pq} .
\end{equation}
The infinitesimal time-step relation (\ref{transmat}) accounts
for the fact that for (infinitesimally) small time-steps $\delta t$ the occupation probabilities change only infinitesimally by $\delta \vec{P}(\delta t) = -\delta t L\cdot \vec{P}(t)$ being of order $\delta t$.
The Laplacian matrix $L$ contains the information on the topological structure of the
network, and acts as the generator matrix of the continuous-time random walk when $\delta t\rightarrow 0$. For undirected networks the Laplacian matrix is
symmetric $L_{pq}=L_{qp}$, a transition matrix defined
by the generating transformation (\ref{transmat}) maintains symmetry at all times $W_{pq}(t)=W_{qp}(t)$.
In view of the structure of the Laplacian matrix (\ref{repres}) we observe that $W(\delta t)=\delta_{pq}(1-\delta tK)+\delta tA_{pq}$.

The adjacency matrix $A_{pq}$ determines the probability transition received by node $p$ from a node $q$ in
a time-step $\delta t$, and $\delta t L_{pp}= K\delta t$ is the amount of probability emitted by a node $p$ to all connected nodes during one-time step $\delta t$.
We see that only connected nodes $p,q$ (with $A_{pq}=1$) can exchange information
during one time-step $\delta t$. We will see subsequently how this limitation of large world networks $K<< N$ changes when instead a fractional power $L^{\frac{\alpha}{2}}$
of the large world network Laplacian matrix $L$ is chosen.

From (\ref{transmat}) follows the interpretation of the Laplacian matrix elements, namely
$L_{pq}$ indicates the probability rate that a node $p$ receives information from node $q$ during a tile-step $\delta t$
and due to $L_{pq}=L_{qp}$ in our model this holds vice versa.
It is important  to chose
$\delta t$ small enough \cite{remark2}
 that  (i) $ W_{pp}(\delta t)= 1-\delta t L_{pp} \geq 0$ where $L_{pp} =K$ meets the requirements to be positive. A further requirement is that all non-vanishing
non-diagonal elements of the Laplacian matrix (\ref{repres}) are negative (ii) $L_{pq} = -A_{pq} < 0 , p\neq q$
so that the elements of the transition matrix are non-negative $0 \leq W_{pq} \leq 1 \forall p \neq q$ having interpretation as probabilities.
It follows then from (i) with
$L_{pp} =-\sum_{p\neq q}L_{pq}= \sum_{p\neq q}A_{pq} $ that the non-diagonal elements additionally fulfill  $W_{pq} = - \delta t L_{pq} < 1$ ($p\neq q$).
We see that the Laplacian of type (\ref{repres})
is meeting this requirement. We come back to this issue in the context of the fractional random walk, as it defines a restriction of admissible range for the exponent
$\alpha$.

For networks where the random walk does not allow the walker to remain at a node $p$ as time increases by an increment
$\delta t=1$, the Laplacian in (\ref{transmat}) has to be renormalized as
${\cal L}_{pq}= \frac{1}{L_{pp}} L_{pq}$ where the diagonal element is the degree of the network $L_{pp}=K_p$
[For networks with variable degrees $K_p$ this leads to non-symmetric transition matrices which is not considered in the present paper].
(For further discussions see e.g. \cite{zhang,NohRieger} and others). As we confine ourselves in the present paper to
regular networks $K_p=K$, this re-normalization of the Laplacian matrix does not change the dynamics as it only rescales the time ($\delta t= \frac{1}{K}$ in (\ref{transmat})).

The time evolution due to (\ref{timemaster}) is then determined by
$\vec{P}(t=m\delta t) = W^m(\delta t)\vec{P}(t=0) =: W(t) \vec{P}(0)$. The matrix element $[W^m(\delta t)]_{pq}=[W(t=m\delta t)]_{pq} $ indicates the probability that the walker starting
at node $q$ reaches after $m$ steps the node $p$.
Let now $\delta t \rightarrow 0$ where $t=m\delta t$ is kept finite then
we get $\delta t$  $\lim_{m\rightarrow \infty} W^m = ({\hat 1}-\frac{t}{m}L)^m = e^{-L t} =W(t)$.
The transition matrix $W(t)$ describes hence the time evolution for the time-continuous random walk
\begin{equation}
\label{timeevol}
P_p(t) = \sum_{q=0}^{N-1} W_{pq}(t)P_q(t=0)  ,\hspace{1cm} W_{pq}(t)= [e^{-L t}]_{pq} .
\end{equation}
The matrix element $W_{pq}$ is the probability of the walker who occupied at $t=0$ node $q$ to reach at time $t$ node $p$. As outlined above the transition matrix of (\ref{timeevol}) is symmetric $W_{pq}(t)=W_{qp}(t)$ for all times $0 \leq t < \infty$.
The transition matrix of the time continuous random walk fulfills hence the evolution equation
\begin{equation}
 \label{transmatcont}
 \frac{dW}{dt} = -L \cdot W(t) ,\hspace{2cm}   \frac{dP_p}{dt} =-\sum_{q=0}^{N-1}L_{pq}P_q \, .
\end{equation}
 Employing Dirac's bra-ket notation the transition matrix has the spectral representation
\begin{equation}
 \label{expo}
 W(t) = e^{-Lt} = |0\rangle\langle 0| +\sum_{\ell =1}^{N-1}e^{-\lambda_{\ell} t}|\ell\rangle\langle \ell|
\end{equation}
where the initial condition
$W(t=0)= {\hat 1} =\sum_{l=0}^{N-1}|\ell\rangle\langle \ell|$ and the stationary distribution (statistic equilibrium)
$W(\infty)= |0\rangle\langle 0|$ with $W_{pq}(\infty)=\frac{1}{N} \forall, p,q$ is rapidly taken.

We refer as `normal random walks' to the random walks described by a Laplacian of the form (\ref{BvK}), in order to distinguish them from 'fractional random walks' which is the subject of the subsequent analysis.

\section{Fractional random walks on regular undirected networks and cubic lattices}

We define the random walk in the same way as in the previous section by performing the transition from time-discrete to time continuous walk.
The only modification is that a fractional power of the network Laplacian occurs as generator matrix for the dynamics. Denoting $W^{(\alpha)}(t)$ the corresponding
'fractional' transition matrix, the time evolution for the fractional continuous random walk is defined by
\begin{equation}
 \label{frac}
 \frac{d}{dt}W^{(\alpha)}(t) = -L^{\frac{\alpha}{2}} \cdot W^{(\alpha)}(t) ,\hspace{2cm} 0<\alpha \leq 2
\end{equation}
where $L$ indicates a Laplacian matrix of type (\ref{repres}).
Now we come back to the crucial conditions (i) and (ii) raised in the previous section on the sign of the fractional Laplacian matrix. It is shown in
the appendix that the fractional Laplacian matrix in the interval $0<\alpha \leq 2$ has negative non-diagonal elements leading to non-negative non-diagonal
elements of the fractional transition matrix.
Hence $0<\alpha\leq 2$ is the
admissible choice for fractional Laplacian matrix $L^{\frac{\alpha}{2}}$ to describe random walk dynamics (\ref{frac}).
This statement on the sign of fractional Laplacian matrix elements is in agreement with previous explicit results for
the periodic chain \cite{michelJphysA}.
So we restrict ourselves on the admissible interval $0<\alpha\leq 2$ for fractional random walks throughout our analysis.
The fractional transition matrix defined by (\ref{frac}) is then again obtained as an exponential having the spectral representation
\begin{equation}
 \label{fracW}
 W^{(\alpha)}(t) = e^{-L^{\frac{\alpha}{2}} t} =  |0 \rangle\langle 0| +\sum_{l=1}^{N-1}e^{-\lambda_{\ell}^{\frac{\alpha}{2}} t}|\ell \rangle\langle \ell| ,\hspace{1cm}
 0 < \alpha \leq 2
\end{equation}
where for $\alpha=2$ the normal random walk transition matrix (\ref{expo}) is recovered. The transition matrix (\ref{fracW}) has the same
stationary transition matrix independent of $\alpha$, namely the equal distribution $W^{(\alpha)}(\infty) =|0 \rangle\langle   0|$
with $W^{(\alpha)}_{pq}(\infty)=\frac{1}{N} \forall p,q$.

Now let us focus on cubic lattices.
The fractional transition matrix (\ref{fracW}) can be written as
\begin{equation}
 \label{Wtransfrac2}
 W^{(\alpha)}_{\vec{p}-\vec{q}}(t) = \frac{1}{N} + \frac{1}{N} \sum_{\vec{\ell} \neq \vec{0}}  e^{i\vec{\kappa}_{\vec{\ell}}\cdot(\vec{p}-\vec{q})}
 e^{-\lambda_{\vec{\ell}}^{\frac{\alpha}{2}} t}\, .
\end{equation}
For large cubic lattices where all $N_j\rightarrow\infty \forall j=1,..,n$ the fractional transition matrix takes the representation\footnote{In the limiting case
$N_j\rightarrow \infty$ the contribution of the stationary limit
$W^{(\alpha)}_{\vec{p}-\vec{q}}(\infty) =\frac{1}{N}=\frac{1}{\prod_{j=1}^nN_j}\rightarrow 0$ can be neglected.}
\begin{equation}
 \label{Wtransfrac}
 \ds W^{(\alpha)}_{\vec{p}-\vec{q}}(t) = \frac{1}{(2\pi)^n}
 \int_{-\pi}^{\pi}{\rm d}\varphi_1 ..\int_{-\pi}^{\pi}{\rm d}\varphi_n \,
 e^{i\sum_{s=1}^n\varphi_s(p_s-q_s)} \, e^{-t\left(2\sum_{j=1}^n(1-\cos{\varphi_j})\right)^{\frac{\alpha}{2}}}
\end{equation}
where the initial condition $W^{(\alpha)}(t=0)={\hat 1}$ is fulfilled by (\ref{Wtransfrac}), namely
$W^{(\alpha)}_{\vec{p}-\vec{q}}(t=0) = \delta_{\vec{p}\vec{q}} = \prod_{j=1}^n\delta_{p_jq_j}$. In (\ref{Wtransfrac}) we introduced $\vec{\varphi} = (\varphi_1, ..,\varphi_n)$ which denotes the quasi-continuous Bloch vector of the infinite
lattice limit [$\varphi_{\ell_s} = \frac{2\pi}{N_s}\ell_s \rightarrow \varphi_s$ and ${\rm d}\varphi_s \sim \frac{2\pi}{N_s}$ as $N_s\rightarrow \infty$ where $-\pi \leq \varphi_s \leq \pi$ can be chosen for the integration limits].

To characterize fractional random walks an interesting question is how the transition matrix behaves asymptotically for large lattices.
Let us first consider the behavior for large times $t>>1$. From (\ref{fracW}) and (\ref{Wtransfrac}) we can see that for large times
the dynamics is dominated by the ensemble of `slowly relaxing' diffusional modes\footnote{The contribution of `fast relaxing modes' is rapidly surpressed.}
$W^{(\alpha)}(t>>1) \sim \sum_{\{\ell_s(t)\}} |\ell_s\rangle\langle\ell_s| e^{-t\lambda_{\ell_s}^{\frac{\alpha}{2}}}$ for which $t\lambda_{\ell_s}^{\frac{\alpha}{2}}$ is still
not large.
It is clear that, the more time $t$ increases, the smaller the set $\{\ell_s(t)\}$ becomes with non-vanishing `surviving modes' corresponding to sufficiently
`small' eigenvalues $\lambda_{\ell_s}^{\frac{\alpha}{2}} << 1$.
 For the cubic lattice the small eigenvalues of the fractional Laplacian are obtained
from (\ref{spectralfracb}) as
$\lambda^{\frac{\alpha}{2}}(\vec{\varphi}) \approx \varphi^{\alpha} $ for sufficiently small $\varphi= |\vec{\varphi}|<< 1$.
For increasingly large times the `surviving modes' are located within an $n$-dimensional
sphere in the $\vec{\varphi}$-wave-vector space (around the origin) where the radius $R$ of this wave-space sphere is developing in time
as $t\lambda_{\ell_s}^{\frac{\alpha}{2}} \leq R^{\alpha}t \sim 1$, i.e. $R_{\alpha}(t)\sim t^{-\frac{1}{\alpha}}$.
The number $D_{\alpha,n}(t)$ of slowly relaxing modes within
this $n$-dimensional sphere `selected to have survived at time $t$' is measured by the volume of this sphere as $D_{\alpha,t}(t) \sim R_{\alpha}^n(t) =
t^{-\frac{n}{\alpha}}$. This means in the large time limit the decay of the fractional transition matrix should behave as
$W^{(\alpha)}_{\vec{p}-\vec{q}}(t)\sim t^{-\frac{n}{\alpha}}$.
We will confirm this argumentation by the subsequent derivation of asymptotic expressions for the fractional transition matrix (\ref{Wtransfrac}).

\subsection{Asymptotic behavior of Fractional transition matrix: Emergence of L\'evy flights}

Let us consider transition matrix (\ref{Wtransfrac}) for large $(\sum_{s=1}^n(p_s-q_s)^2)^{\frac{1}{2}} = |\vec{p}-\vec{q}| >>1$ on large lattices and finite times.
We then can evaluate (\ref{Wtransfrac2})
by introducing the vector $\vec{k}$ with the components $k_j = \varphi_j|\vec{p}-\vec{q}| $ where the integration limits can be put $\pm \infty$ to obtain the leading asymptotic contribution
\begin{equation}
 \label{Wtransfrac3}
 W^{(\alpha)}_{\vec{p}-\vec{q}}(t) \approx \frac{1}{(2\pi|\vec{p}-\vec{q}|)^n}
 \int_{-\infty}^{\infty}{\rm d}k_1 ..\int_{-\infty}^{\infty}{\rm d}k_n \, e^{i\vec{k}\cdot\frac{(\vec{p}-\vec{q})}{|\vec{p}-\vec{q}|}} \,
 e^{-t\left(2\sum_{j=1}^n\left(1-\cos{\frac{k_j}{|\vec{p}-\vec{q}|}}\right)\right)^{\frac{\alpha}{2}}} .
\end{equation}
Let us consider now this integral for large $|\vec{p}-\vec{q}|^{\alpha} >> t $ where asymptotically this integral is dominated by $\left(2\sum_{j=1}^n\left(1-\cos{\frac{k_j}{|\vec{p}-\vec{q}|}}\right)\right)^{\frac{\alpha}{2}} \sim
 \frac{|\vec{k}|^{\alpha}}{|\vec{p}-\vec{q}|^{\alpha}} $ to arrive at the asymptotic expression
 \begin{equation}
 \label{asym}
 W^{(\alpha)}_{\vec{p}-\vec{q}}(t) \approx \frac{1}{(2\pi)^n}
 \int_{-\infty}^{\infty}{\rm d}k_1 ..\int_{-\infty}^{\infty}{\rm d}k_n \, e^{i\vec{k}\cdot(\vec{p}-\vec{q})} e^{-t|\vec{k}|^{\alpha}} .
\end{equation}
This asymptotic representation for the fractional transition matrix is within the admissible interval $0<\alpha \leq 2$,
the well known Fourier representation of an $\alpha$-stable L\'evy distribution in the $n$-dimensional infinite space, e.g. \cite{Chechkin2006}.
The cases $\alpha =1$ and $\alpha =2$ indicate Cauchy- and normal distributions,
respectively.

Having arrived at the asymptotic relation (\ref{asym}) for the fractional transition matrix, it is straight-forward to obtain the representation
\footnote{Put in (\ref{asym}) $\vec{k}=\vec{\xi}t^{-\frac{1}{\alpha}}$.} \cite{michel-et-al-13},
and see also \cite{metzler}
\begin{equation}
\label{asy2}
 W^{(\alpha)}_{\vec{p}-\vec{q}}(t) \approx t^{-\frac{n}{\alpha}} {\tilde W}\left(\frac{|\vec{p}-\vec{q}|^{\alpha}}{t}\right)
\end{equation}
where function ${\tilde W}$ depends only on the ratio $\frac{|\vec{p}-\vec{q}|^{\alpha}}{t}$ and fulfills
\begin{equation}
\label{wfu}
\begin{array}{l}
{\tilde W}(\frac{|\vec{p}-\vec{q}|^{\alpha}}{t}\rightarrow 0) = \frac{1}{(2\pi)^n}\int {\rm d}^n\vec{\xi} e^{-\xi^{\alpha}}  \\ \\
{\tilde W}(0) = \frac{1}{(2\pi)^n} \frac{2\pi^{\frac{n}{2}}}{\Gamma(\frac{n}{2})}\int_0^{\infty} e^{-\xi^{\alpha}}\xi^{n-1}{\rm d}\xi =
\frac{2}{(2\sqrt{\pi})^n}\frac{2}{\alpha} \frac{\Gamma (\frac{n}{\alpha})}{\Gamma (\frac{n}{2})}
\end{array}
\end{equation}
and finally, the asymptotic expression for $W^{(\alpha)}_{\vec{p}-\vec{q}}(t)$ for large lattices and $t\gg 1$ is given by:
\begin{equation}
\label{Wasyt}
W^{(\alpha)}_{\vec{p}-\vec{q}}(t) \approx \frac{2}{(2\sqrt{\pi})^n}\frac{2}{\alpha} \frac{\Gamma (\frac{n}{\alpha})}{\Gamma (\frac{n}{2})}t^{-\frac{n}{\alpha}}.
\end{equation}
The result in equation (\ref{Wasyt}) establishes the probabilities of transition at time $t$ from the stating position $\vec{q}$ to the final location determined
by $\vec{p}$ for a random walker following the fractional dynamics determined by equation (\ref{frac}); this result is discussed in the following part in the context of probabilities of return.
\\[2mm]
On the other hand, the transition probabilities at each step are given by the transition matrix:
\begin{equation}
\label{Wlattice}
W^{(\alpha)}_{\vec{p}-\vec{q}}(t) \approx \frac{1}{(2\pi)^n}
 \int_{-\infty}^{\infty}{\rm d}k_1 ..\int_{-\infty}^{\infty}{\rm d}k_n \, e^{i\vec{k}\cdot(\vec{p}-\vec{q})} e^{-|\vec{k}|^{\alpha}t}
\end{equation}
and a similar approach allows to obtain for $|\vec{p}-\vec{q}|^{\alpha} >> tC_{\alpha,n}$:
\begin{equation}
\label{Wlattice_limit}
W^{(\alpha)}_{\vec{p}-\vec{q}} \propto \frac{tC_{\alpha,n}}{|\vec{p}-\vec{q}|^{n+\alpha}}
\end{equation}
where the constant $C_{\alpha,n}=\frac{2^{\alpha-1}\alpha\Gamma(\frac{\alpha+n}{2})}{\pi^{\frac{n}{2}}\Gamma(1-\frac{\alpha}{2})}$
is given in the literature \cite{michel-et-al-13,michel-et-al-14} (having here physical dimension [sec$^{-1}$]).
The result in Eq. (\ref{Wlattice_limit}) for $n$-dimensional lattices emerges from the fractional Laplacian ${\cal L}^{\frac{\alpha}{2}}$. This power-law decay $|\vec{p}-\vec{q}|^{-n-\alpha}$ for the transition probabilities of a random walker is the footprint of a dynamics with L\'evy flights in $n$-dimensions defined within admissible interval $0<\alpha \leq 2$.
\\[2mm]
The time derivative of (\ref{asym}) yields
\begin{equation}
 \label{Wtransfrac4}
 \begin{array}{l}
 \ds \frac{d}{dt} W^{(\alpha)}_{\vec{p}-\vec{q}}(t=0) = -{\cal L}^{\frac{\alpha}{2}}_{\vec{p}-\vec{q}} \approx \frac{C_{\alpha,n}}{|\vec{p}-\vec{q}|^{n+\alpha}} \\ \\ \hspace{0.5cm}
\ds = \frac{1}{|\vec{p}-\vec{q}|^{n+\alpha}} \frac{(-1)}{(2\pi)^n}
 \int_{-\infty}^{\infty}{\rm d}k_1 ..\int_{-\infty}^{\infty}{\rm d}k_n \, e^{i\vec{k}\cdot\frac{(\vec{p}-\vec{q})}{|\vec{p}-\vec{q}|}}|\vec{k}|^{\alpha} \, \\ \\ \hspace{0.5cm}
 \ds = \frac{(-1)}{(2\pi)^n}
 \int_{-\infty}^{\infty}{\rm d}k_1 ..\int_{-\infty}^{\infty}{\rm d}k_n \, e^{i\vec{k}\cdot(\vec{p}-\vec{q})} |\vec{k}|^{\alpha} \\ \\ \hspace{0.5cm}
 \ds \frac{d}{dt} W^{(\alpha)}_{\vec{p}-\vec{q}}(t=0)  \approx - (-\Delta_n)^{\frac{\alpha}{2}}\delta^{n}(\vec{p}-\vec{q}) .
 \end{array}
\end{equation}
The last relation is nothing but the fractional diffusion equation of a L\'evy flyer in the $n$-dimensional continuous space where the fractional
Laplacian matrix takes the representation of a Riesz fractional derivative kernel (continuous fractional Laplacian kernel)  \cite{michel-et-al-13,riesz2}.
%
%
%%%%%%%%%%%%%%%%%%%%%%%%%%%%%%%%%%%%%%%%%%%%%%%%%%%%%%%%
%
%%%%%%%%%%%%%%%%%%%%%%%%%%%%%%%%%%%%%%%%%%%%%%%%%%%%%%%%
%

\subsection{Fractional return probabilities}

Now closely connected with the derivations of the previous paragraph, let us analyze
the return probability of the fractional random walk which is defined as the probability that a walker starting at $t=0$ at a node $p$ returns at this node
after time $t$. The return probability is obtained by the uniform value of the diagonal elements of the transition matrix $W^{(\alpha)}(t)$, namely
\begin{equation}
\label{returnpro}
p_{\alpha}(t) = W^{(\alpha)}_{pp}(t) = \frac{1}{N}trace(W^{(\alpha)}(t)) =\frac{1}{N}+ \frac{1}{N} \sum_{l=1}^{N-1}e^{-\lambda_l^{\frac{\alpha}{2}} t}
\end{equation}
having identical values for all nodes $p$,
decaying rapidly to the stationary equal distribution $p(\infty)=\frac{1}{N}$ and obeying the initial condition $p_{\alpha}(t=0) = 1$. Since all diagonal elements
of the transition matrix have the same value, the return probability for any node $p$ coincides with the global average return probability.

Let us now consider large cubic lattices  $N_j\rightarrow\infty$, $ \forall j=1,..,n$. Then the return probability is obtained by
the diagonal elements of (\ref{Wtransfrac}) as
[where $p_{\alpha}(\infty)=\frac{1}{N} \rightarrow 0$ can be set to zero]
\begin{equation}
 \label{inreturnprob}
 p_{\alpha}(t)-p_{\alpha}(\infty) = \left(W^{(\alpha)}_{\vec{0}}(t)-W^{(\alpha)}_{\vec{0}}(\infty)\right) =
 \frac{1}{(2\pi)^n}
 \int_{-\pi}^{\pi}{\rm d}\varphi_1 ..\int_{-\pi}^{\pi}{\rm d}\varphi_n e^{-t\left(2\sum_{j=1}^n(1-\cos{\varphi_j})\right)^{\frac{\alpha}{2}}} .
\end{equation}
The return probability of the normal random walk is recovered by (\ref{inreturnprob}) for $\alpha=2$ and yields $p_2(t) = e^{-2n t} I_0^n(2t)$ in accordance with the well known
expression of the literature $ e^{-t} I_0^n(\frac{t}{n})$ which is recovered \cite{riascos-fracdyn} when instead of the Laplacian matrix $L_{pq}$,
the modified Laplacian matrix $ \frac{L_{pq}}{K}=\frac{1}{2n}{\cal L}_{pq}$ is employed. The large time asymptotics of the return probability can be obtained
from expression by setting $|\vec{p}-\vec{q}|=0$ in the asymptotic relation (\ref{asy2}) and accounting for (\ref{wfu}) we arrive at

\begin{equation}
 \label{expli}
 p_{\alpha}(t) - p_{\alpha}(\infty) \approx \frac{2}{(2\sqrt{\pi})^n}\frac{2}{\alpha} \frac{\Gamma (\frac{n}{\alpha})}{\Gamma (\frac{n}{2})} \, t^{-\frac{n}{\alpha}} .
\end{equation}
The fractional return probability decays for large cubic lattices in $n$-dimensions also
with a characteristic $t^{-\frac{n}{\alpha}}$-power law.
We have to interpret this power law decay in the same way as for the fractional transition matrix, namely as a collective effect of such normal modes with Bloch-wave vectors $\vec{\kappa}$ located on a in the limit infinitesimal
small sphere around
the origin ${\vec{\kappa}} = 0$ where in this region $\lambda_{\vec{\ell}}\rightarrow 0$. The large time power law behavior emerges only in sufficiently
large lattices, i.e. when the infinitesimal small $\vec{\kappa}$-sphere is densely populated with normal modes.
The collective effect
of these long wave diffusional modes act like a L\'evy flyer in the continuous space where the discrete structure of
the lattice is ignored.
For the cyclic ring $n=1$ the $ p_{\alpha}(t) - p_{\alpha}(\infty) \sim t^{-\frac{1}{\alpha}}$ given in \cite{riascos-fracdyn} is recovered as well as the
$p_{\alpha}(t) - p_{\alpha}(\infty)  \sim t^{-\frac{n}{2}}$ for normal random walks
($\alpha=2$) on the $n$-dimensional cubic lattice. We notice that in range $0<\alpha< 2$, the relaxation of the fractional random walk
$p_{\alpha}(t) - p_{\alpha}(\infty) \sim t^{-\frac{n}{\alpha}}$ is faster than those of the normal random walk $\alpha=2$ in the same lattice.
In the extreme limit of vanishing (infinitesimally positive) $\alpha\rightarrow 0$, it can be seen that the return probability $p_{\alpha}(t)$ starting at
$p_{\alpha}(t=0)=1$ is instantaneously collapsing as $t^{-\frac{n}{\alpha}}$ versus the equilibrium value $p_{\alpha}(\infty)=\frac{1}{N}$.
From this observation follows that the fractional walk on the $n$-dimensional becomes the faster, the smaller $\alpha$.

\subsection{Derivation of the fractional fundamental matrix}

The goal of this paragraph is to obtain further characteristic features of the fractional random walk such as for instance mean relaxation times which
also is referred to as `global time' \cite{riascos-fracdyn}
and closely related the Kemeny constant. We demonstrate that this information is contained in the fractional
 Green's matrix to be deduced in this paragraph, and closely related, the fractional fundamental matrix.
To this end we
note that $L^{\frac{\alpha}{2}}$ due to its eigenvalue $\lambda_0=0$ is not invertible in the $N$-dimensional space of all $N$ eigenvectors $| \ell\rangle $. However it is
invertible within the $N-1$-dimensional subspace of eigenvectors $|\ell \rangle$ of positive eigenvalues, i.e. the basis of this subspace does not contain the eigenvector $|0\rangle$
with $L|0\rangle = 0$. Denoting the representation of the Laplacian matrix in
this subspace as ${\tilde L}$ then  we have
\begin{equation}
 \label{fracWsub}
 W^{(\alpha)}(t)-W^{(\alpha)}(\infty) = e^{-{\tilde L}^{\frac{\alpha}{2}} t} =\sum_{\ell=1}^{N-1}e^{-\lambda_{\ell}^{\frac{\alpha}{2}} t}|\ell\rangle\langle\ell|
\end{equation}
where the spectral sum contains only relaxing terms, where
${\tilde L}^{\frac{\alpha}{2}}$ is invertible in this $N-1$-dimensional subspace [with unity ${\hat 1}_{(N-1)\times (N-1)}=
{\hat 1}_{N\times N}-|0\rangle\langle 0| =\sum_{l=1}^{N-1}|\ell\rangle\langle\ell |$]
\begin{equation}
 \label{invert}
 {\tilde L}^{-\frac{\alpha}{2}} = \int_0^{\infty} (W^{(\alpha)}(t)-W^{(\alpha)}(\infty)){\rm d}t =
\sum_{\ell=1}^{N-1}\lambda_{\ell}^{-\frac{\alpha}{2}}|\ell\rangle\langle\ell|
\end{equation}
We refer to as (\ref{invert}) the fractional Green's matrix.
We see that the set of eigenvalues of the (fractional) Green's matrix (\ref{invert}) is the set of relaxation times
$\tau_{\ell}^{\alpha} = \lambda_{\ell}^{-\frac{\alpha}{2}}$ occuring in (\ref{fracWsub}).
To link the Green's matrix with some common notions in the literature, let us evoke the definition of the so called `fundamental matrix' $Z$ which is defined in
in the literature for time-discrete random walks as \cite{kemeny,zhang}
\begin{equation}
\label{relaxing}
Z = \delta t \sum_{m=0}^{\infty} [W(\delta t)-W(\infty)]^m= \delta t [{\hat 1}-(W(\delta t)-W(\infty))]^{-1}
\end{equation}
where $W(\delta t=1)$ is the transition matrix corresponding to one time-step (\ref{transmat})
which is converging since all $N-1$ eigenvalues $(1- \delta t \lambda_{\ell}^{-\frac{\alpha}{2}})$ of $[W(\delta t)-W(\infty)]$ are inferior to $1$. In contrast to \cite{zhang}
where the time-step is chosen $\delta t = 1$, we introduce here the transition matrix
with a multiplyer of the time increment $\delta t$, so that $Z$ has physical dimension of time. In this way we can account for time-discrete, and in the limit
$\delta t \rightarrow 0$, for time-continuous random walks.

For the time-discrete fractional random walk, we employ in (\ref{relaxing})
$W^{(\alpha)}(\delta t)-W^{(\alpha)}(\infty)= {\hat 1}-|0\rangle\langle0|-\delta t L^{\frac{\alpha}{2}} = {\hat 1}_{N-1\times N-1}-\delta t {\tilde L}^{\frac{\alpha}{2}} $ which
is just the relaxing part having spectral representation $W^{(\alpha)}(\delta t)-W^{(\alpha)}(\infty)= \sum_{\ell=1}^{N-1}|\ell\rangle\langle\ell|
(1-\delta t \lambda_{\ell}^{\frac{\alpha}{2}})$. Now in order to evaluate the geometrical series
(\ref{relaxing}) we have to pay attention on the zero order  $[W(\delta t)-W(\infty)]^0={\hat 1}_{N\times N} = |0\rangle\langle 0|+{\hat 1}_{(N-1) \times (N-1)}$ being the complete $N\times N$-unity matrix which containing
$0\rangle\langle 0|$ which is not in the $N-1$-dimensional subspace of relaxing modes, whereas all orders $m>0$
have only contributions within in the $N-1$-dimensional subspace.
We can hence write for the expansion of the fractional fundamental matrix\footnote{As $|[(1-\delta t \lambda_{\ell}^{\frac{\alpha}{2}})| < 1$
this is a converging geometrical series $\sum_{m=0}^{\infty}[(1-\delta t \lambda_{\ell}^{\frac{\alpha}{2}})]^m = [1-(1-\delta t \lambda_{\ell}^{\frac{\alpha}{2}})] ^{-1}=
(\delta t)^{-1}\lambda_{\ell}^{-\frac{\alpha}{2}}$  as $|[(1-\delta t \lambda_{\ell}^{\frac{\alpha}{2}})| < 1$ $\forall \ell=1,..,N-1$.}
\begin{equation}
\label{relaxfrac}
\begin{array}{l}
\ds Z^{(\alpha)} = \delta t \sum_{m=0}^{\infty} [W^{(\alpha)}(\delta t)-W(\infty)]^m=\delta t
[1- (W^{(\alpha)}(\delta t)-W(\infty) ) ]^{-1} \\ \\ \ds Z^{(\alpha)}  =\delta t\left(|0\rangle\langle 0|+  \sum_{\ell=1}^{N-1}|\ell\rangle\langle\ell|
\sum_{m=0}^{\infty}[(1-\delta t \lambda_{\ell}^{\frac{\alpha}{2}})]^m\right)  \\ \\ \ds Z^{(\alpha)}  = \delta t |0\rangle\langle 0|+ \sum_{\ell=1}^{N-1}|\ell\rangle\langle\ell|
\lambda_{\ell}^{-\frac{\alpha}{2}} = \delta t |0\rangle\langle 0| + {\tilde L}^{-\frac{\alpha}{2}} .
\end{array}
\end{equation}
We see that the fractional fundamental matrix $Z^{(\alpha)}$ for a discrete time walk is related to our above
introduced fractional Green's function for the time-continuous walk (\ref{invert}) and coinciding in the limit $\delta t\rightarrow 0$.
It is clear that this reasoning remains true for any transition from time-discrete to time-continuous random walks, independent of the choice of Laplacian matrix.
Relation (\ref{relaxfrac}) is consistent with spectral representations given in the literature for discrete time random walks \cite{zhang} [where there $\delta t=1$].
Indeed the fractional fundamental matrix $Z^{(\alpha)}$ contains the entire statistical information needed to characterize the `efficiency' of fractional random walk.

Before we exploit this information more closely for the cubic lattice, let us derive some important analytical relations between $Z^{(\alpha)}$ and
normal random walk transition matrix $W^{(2)}$.
To this end account first for the following integral relation holding for scalar $\lambda >0$ and $\alpha >0$ [where we confine our analysis nevertheless as emphasized above
to the admissible range
$0<\alpha\leq 2$]
\begin{equation}
 \label{intrel}
 \lambda^{-\frac{\alpha}{2}} = \frac{1}{\Gamma(\frac{\alpha}{2})}\int_0^{\infty}e^{-\lambda t} t^{\frac{\alpha}{2}-1}{\rm d} t .
\end{equation}
By using this relation and the spectral representation (\ref{relaxfrac}) we can represent $Z^{(\alpha)}$ for the time-continuous walk by
\begin{equation}
\label{fowa}
\begin{array}{l}
\ds  Z^{(\alpha)} =   {\tilde L}^{-\frac{\alpha}{2}} = \frac{1}{\Gamma(\frac{\alpha}{2})}\int_0^{\infty}e^{-{\tilde L}t} \, t^{\frac{\alpha}{2}-1}{\rm d} t =
 \frac{1}{\Gamma(\frac{\alpha}{2})}\int_0^{\infty}(W^{(2)}(t)-W^{(2)}(\infty)) t^{\frac{\alpha}{2}-1}{\rm d} t  \\ \\
 \ds \hspace{0.5cm} = \int_0^{\infty} e^{- {\tilde L}^{\frac{\alpha}{2}} t}{\rm d}t =
 \int_0^{\infty}(W^{(\alpha)}(t)-W^{(\alpha)}(\infty)){\rm d} t .
 \end{array}
\end{equation}
The relations (\ref{intrel}), (\ref{fowa}) hold in the entire admissible range $0<\alpha \leq 2$
where $W^{(2)}(t)= e^{-Lt}$ is the transition matrix of the normal random walk. The representation (\ref{fowa}) is especially useful when the normal walk
transition matrix is explicitly known.
As we will see, this is the case for normal walks on cubic $n$-dimensional lattices.
Let us now evaluate (\ref{fowa}) to obtain the fractional fundamental matrix of cubic $n$-dimensional infinite lattice where all $N_j\rightarrow\infty$.
Then in view of (\ref{spectralfracb}) the spectral sums take integral representations $\frac{1}{N}\sum_{\vec{\ell}}(..) \rightarrow \frac{1}{(2\pi)^n} \int_{-\pi}^{\pi}...\int_{-\pi}^{\pi}(..){\rm d}\varphi_1..{\rm d}{\varphi_n}$.
The transition matrix for the normal walk which is obtained from (\ref{Wtransfrac}) for $\alpha=2$ and yields
\begin{equation}
 \label{normalwalk}
 W_{\vec{p}-\vec{q}}^{(2)}(t) = \frac{1}{(2\pi)^n}
 \int_{-\pi}^{\pi}{\rm d}\varphi_1 ..\int_{-\pi}^{\pi}{\rm d}\varphi_n e^{-t\left(2\sum_{j=1}^n(1-\cos{\varphi_j})\right)}e^{i(\vec{p}-\vec{q})\cdot{\vec{\varphi}}}
= e^{-2n t} \prod_{j=1}^n I_{|p_j-q_j|}(2t)
\end{equation}
where $I_p(\tau)=\frac{1}{\pi}\int_0^{\pi}e^{\tau\cos{\varphi}}\cos{(p\varphi)}{\rm d}\varphi $ denotes the modified Bessel function of the first kind of integer orders $p$
\cite{abramo}. Using (\ref{fowa}) we can determine the fractional fundamental matrix for infinite cubic lattices as
\begin{equation}
\label{fowafrac}
Z^{(\alpha)}_{p_1-q_1,..,p_j-q_j,..,p_n-q_n}
= \frac{1}{\Gamma(\frac{\alpha}{2})}\int_0^{\infty} e^{-2n t}  t^{\frac{\alpha}{2}-1} \prod_{j=1}^n I_{|p_j-q_j|}(2t) {\rm d} t  =
\frac{1}{(2n)^{\frac{\alpha}{2}}\Gamma(\frac{\alpha}{2})}
\int_0^{\infty} \tau^{\frac{\alpha}{2}-1}  e^{-\tau}\prod_{j=1}^n I_{|p_j-q_j|}\left(\frac{\tau}{n}\right)  {\rm d}t
\end{equation}
containing the normal random walk transition matrix elements \\ $W^{(2)}_{p_1-q_1,..,p_j-q_j,..,p_n-q_n}\left(\frac{\tau}{2n}\right) =
e^{-\tau}\prod_{j=1}^n I_{|p_j-q_j|}\left(\frac{\tau}{n}\right)$ (in rescaled time), in accordance with the known expression $e^{-\tau}I_{|p-q|}(\tau)$ for $n=1$ \cite{redner}.

\subsection{Fractional mean relaxation time}

Having arrived at the representation (\ref{fowafrac}) for the fractional fundamental matrix of the cubic lattice, it is now not a big deal to determine the
fractional mean relaxation time ${\bar \tau}_{\alpha} = \frac{1}{N}trace(Z^{(\alpha)})=\frac{1}{N}\sum_{\ell=1}^{N-1}\lambda_{\ell}^{-\frac{\alpha}{2}} $
(for a further  discussion see \cite{doyle,riascos-fracdyn,riascos-fracdiff,zhang}).
The smaller mean relaxation time, the faster the stationary stationary distribution is reached, and the higher the efficiency
for the walker to explore the network.

From the above considerations follows that the fractional global time ${\bar \tau}_{\alpha}$ is determined by the trace of the fractional fundamental matrix (\ref{relaxfrac}),
and for the infinite cubic lattice by (\ref{fowafrac}). The fractional mean relaxation time of the cubic lattice is then the uniform value of the diagonal elements of (\ref{fowafrac})

\begin{equation}
\label{fowafractrace}
\begin{array}{l}
\ds {\bar \tau}_{\alpha} =Z^{(\alpha)}_{0,..,0}= \frac{1}{\Gamma(\frac{\alpha}{2})}\int_0^{\infty} \left(W^{(2)}_{\vec{0}}(t)-W^{(2)}_{\vec{0}}(\infty)\right) t^{\frac{\alpha}{2}-1}
{\rm d} t \\ \\  =
\ds \frac{1}{\Gamma(\frac{\alpha}{2})}\int_0^{\infty} e^{-2n t} \left(I_{0}(2t)\right)^n t^{\frac{\alpha}{2}-1} {\rm d} t  \\  \\ =
\ds \frac{1}{(2n)^{\frac{\alpha}{2}}\Gamma(\frac{\alpha}{2}) }\ds \int_0^{\infty} e^{-\tau}
\left(I_0\left(\frac{\tau}{n}\right)\right)^n \, \tau^{\frac{\alpha}{2}-1}  {\rm d}\tau .
\end{array}
\end{equation}
Interesting is the limiting case when $\alpha\rightarrow 0+$ infinitesimally positive, then since
$\frac{1}{\Gamma(\frac{\alpha}{2})}\int_0^{\infty} e^{-\tau} \, \tau^{\frac{\alpha}{2}-1}  {\rm d}\tau =1 $ is a normalized integral which is for $\alpha\rightarrow 0+$ concentrated
arround $\tau=0$, we can write for (\ref{fowafractrace}) in this limiting case
\begin{equation}
 \label{limit}
  {\bar \tau}_{0+} =\lim_{\alpha\rightarrow 0+}{\bar \tau}_{\alpha}  =  (I_0(0))^n \frac{1}{\Gamma(\frac{\alpha}{2})}\int_0^{\infty} e^{-\tau} \, \tau^{\frac{\alpha}{2}-1}  {\rm d}\tau =1
\end{equation}
since $I_0(0)= 1$.
The limiting value ${\bar \tau}_{\alpha \rightarrow 0+}$ is in accordance with the observation that yields the limit of the diagonal element of (\ref{invert}), namely
${\bar \tau}_{\alpha \rightarrow 0+} = \lim_{\alpha\rightarrow 0} \frac{1}{N}\sum_{\ell=1}^{N-1}\lambda_{\ell}^{-\frac{\alpha}{2}} =\frac{N-1}{N} \rightarrow 1$ for large lattices.

\vskip0.5cm
\begin{center}
\includegraphics[width=0.6\textwidth]{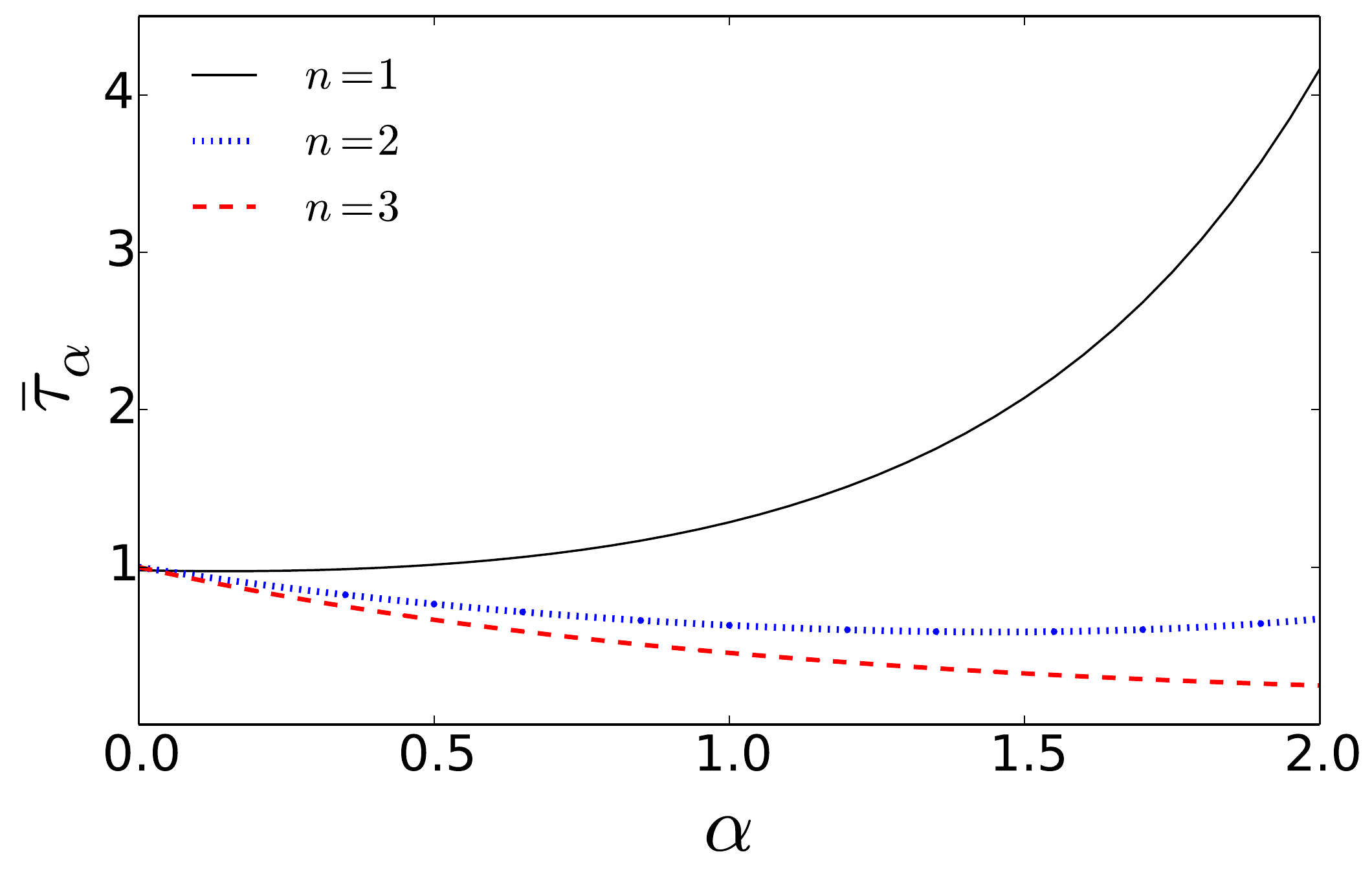}
\end{center}
\vspace{-7mm}
{\sf Fig. 1: Representation of the fractional characteristic time (mean relaxation time) ${\bar \tau}_{\alpha} = \frac{1}{N}\sum_{\ell=1}^{N-1}\lambda_{\ell}^{-\frac{\alpha}{2}}$ as function of $\alpha$ for the cubic lattice with $N= 50^n$  nodes. We depict the results obtained for $n=1,2,3$.}
\newline\newline

Figure 1 shows the mean relaxation
time ${\bar \tau}_{\alpha} = \frac{1}{N}\sum_{\ell=1}^{N-1}\lambda_{\ell}^{-\frac{\alpha}{2}}$ for the cubic lattice with $N= 50^n$ nodes for dimensions $n=1,2,3$, respectively.
We see that the mean relaxation time
 in the fractional range $0 <\alpha < 2$ is in the one-dimensional case $n=1$ reduced compared with the normal walk $\alpha=2$.
For one-dimensional cyclic chains the relaxation to the stationary distribution is faster for the fractional walk
as for the normal walk.
We demonstrate in the subsequent subsection that a fractional random walk dynamics governed by a renormalized fractional Laplacian matrix
enhances the speed of fractional random walk. In the following paragraph we compare the efficiency of fractional random walks as a function of $\alpha$. The appropriate measure to do
so is to consider the Kemeny constants of these walks. This is outlined in the subsequent part.

\subsection{Efficiency of fractional random walk}

In order to compare efficiency of fractional random walks for different $\alpha$ let us consider
again the time-discrete fractional random walk which is generated by (where we put here without loss of generality $\delta t=1$)
\begin{equation}
 \label{modi}
 \vec{P}^{(\alpha)}(t+1)= ({\hat 1}-{\tilde {\cal L}}^{(\alpha)}) \cdot \vec{P}^{(\alpha)} (t) = \frac{1}{k^{(\alpha)}}A^{(\alpha)}\cdot \vec{P}^{\alpha)}(t) .
\end{equation}
In order to compare the dynamics for different $\alpha$ we need to consider fractional random walks for different $\alpha$ on the same time-scale.
This is achieved when we normalize the fractional Laplacian matrices in (\ref{modi}) such that they all have the same diagonal elements ${\tilde {\cal L}}^{(\alpha)}_{pp}=1$
in the order of magnitude of the time step, i.e. independent of $\alpha$. Hence we employ in the master equation (\ref{modi}) the re-normalized fractional Laplacian matrix

\begin{equation}
\label{modifieddef}
{\tilde {\cal L}}^{(\alpha)}_{pq} = \frac{1}{k^{(\alpha)}}L^{\frac{\alpha}{2}}_{pq} = {\hat 1} - \frac{1}{k^{(\alpha)}}A^{(\alpha)}_{pq} ,
\end{equation}
where  $k^{(\alpha)}= L^{\frac{\alpha}{2}}_{pp} =\frac{1}{N}trace(L^{\frac{\alpha}{2}})=\frac{1}{N}\sum_{\ell=1}^{N-1}\lambda_{\ell}^{\frac{\alpha}{2}}$
denotes the fractional degree and $A^{(\alpha)}_{pq}$ the fractional adjacency matrix
which is constructed by the negative non-diagonal elements of the fractional Laplacian matrix $L^{\frac{\alpha}{2}}$ of the type (\ref{repres})
(i.e. $A^{(\alpha)}_{pp}=0$ and $A^{(\alpha)}_{pq}= -L^{\frac{\alpha}{2}}_{pq}$ for $p\neq q$)
\cite{riascos-fracdyn,riascos-fracdiff}. In the appendix we have shown that $0<\alpha\leq 2$ is the admissible interval where the fractional Laplacian
has required good properties that the fractional connectivity matrix has positive non-vanishing elements $A^{(\alpha)}_{pq} \geq 0$.
In view of (\ref{modi}) the quantities $\frac{1}{k^{(\alpha)}}A^{(\alpha)}$ denote the probabilities that
the walker located at node $q$ moves to node $p$ where because of $A^{(\alpha)}_{pp}=0$ the walker must move to another node as time increases
by a time-step $\delta t=1$. This is reflected
by the fact the normalization condition $\sum_{p=0}^{N-1}\frac{1}{k^{(\alpha)}}A^{(\alpha)}_{pq} =1$ and by $0 < A^{(\alpha)}_{pq} <1$ within the admissible range $0<\alpha\leq 2$.
For the time-discrete fractional random walk characterized by (\ref{modi}), the eigenvalues of ${\cal L}^{(\alpha)}$ take the form
$\frac{\lambda_{\ell}^{\frac{\alpha}{2}}}{k^{(\alpha)}}$ [where $\lambda_{\ell}$ denote as above the eigenvalues of (\ref{repres})].

\subsubsection{Fractional Kemeny constant}

The Kemeny constant for this type of random walk is with (\ref{relaxfrac}) determined by \cite{kemeny}\footnote{where in the literature the definition of Kemeny constant often
differs by the constant $1$.}

\begin{equation}
 \label{kemenymodi}
 {\mathcal K}^{(\alpha)} = trace\left[\left(\frac{{\tilde L}^{\frac{\alpha}{2}}}{k^{(\alpha)}}\right)^{-1}\right]  =  k^{(\alpha)}\sum_{\ell=1}^{N-1}\lambda_{\ell}^{-\frac{\alpha}{2}}
\end{equation}
where $\left(\frac{{\tilde L}^{\frac{\alpha}{2}}}{k^{(\alpha)}}\right)^{-1}$ denotes the modified fractional
Green's function occuring in the fractional fundamental matrix (\ref{relaxfrac}) when the renormalized fractional Laplacian (\ref{modifieddef}) is employed.
In the study of discrete time random walks, the Kemeny constant is a global time that, for regular networks,
gives the average of the times needed by the random walker to reach any site of the structure; this value can be calculated in terms of the
Laplacian eigenvalues \cite{riascos-fracdiff}. [For further derivations and definitions of Kemeny constant
we refer to \cite{doyle,kemeny}].
For the fractional dynamics on $n$-dimensional lattices, the Kemeny $\mathcal{K}^{(\alpha)}$ constant is given by:
\begin{equation}
\mathcal{K}^{(\alpha)}\equiv\left(\frac{1}{N}\sum_{m=1}^{N-1} \lambda_{m}^{\alpha/2}\right)\left(\sum_{l=1}^{N-1} \lambda_{l}^{-\alpha/2}\right).
\end{equation}
The value $k^{(\alpha)}\equiv\frac{1}{N}\sum_{m=1}^{N-1} \lambda_{m}^{\alpha/2}=\frac{1}{N}trace(L^{\alpha/2})$ is the fractional degree.
For $n$-dimensional lattices with $N\gg 1$ is $k^{(\alpha)}\approx (2n)^{\alpha/2}$.
The Kemeny constant for lattices in $n=1,2,3$ dimensions is depicted in Figure 2. We see how this characteristic time, in the fractional range $0 <\alpha < 2$ is reduced, compared with the normal random walk $\alpha=2$.
The relaxation to the stationary distribution is faster for the fractional walk
as for the normal walk. A fractional random walk search strategy to explore the lattice has thus due to long range moves and emerging a small-world property
increasing the capacity of the random walker to reach any site of the lattice in comparison with the normal strategy.
These results confirm the recent findings in \cite{riascos-fracdyn,riascos-fracdiff}.

\vskip0.5cm
\begin{center}
\includegraphics[width=0.6\textwidth]{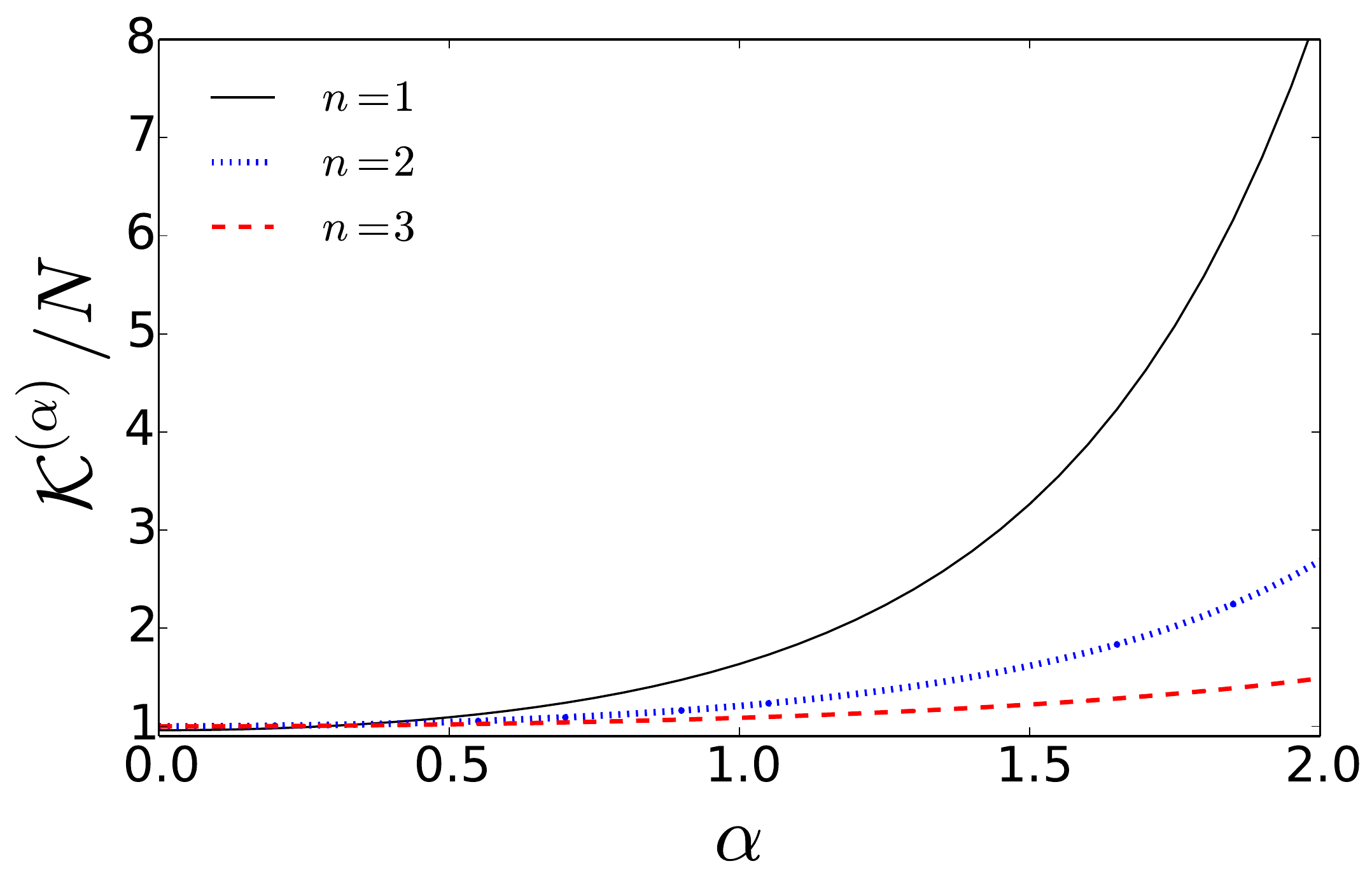}
\end{center}
\vspace{-7mm}
{\sf  Fig. 2:  Representation of the Kemeny constant  $\mathcal{K}^{(\alpha)}$ as function of $\alpha$ for the cubic lattice with $N= 50^n$  nodes. We depict the results obtained for $n=1,2,3$.
}
\newline

In Figure 2 we plot $\mathcal{K}^{(\alpha)} =\frac{1}{N}\sum_{m=1}^{N-1}\lambda_{m}^{\frac{\alpha}{2}}\, \sum_{\ell=1}^{N-1}\lambda_{\ell}^{-\frac{\alpha}{2}}  $ for the infinite cubic lattice for spatial dimensions $n=1,2,3$.
This figure shows that the random walk governed by (\ref{modi}) is always faster the smaller $\alpha$ in $0<\alpha < 2$ for
a fractional random walk, as compared to a normal random walk $\alpha=2$. A fractional random walk search strategy
to explore the lattice has thus due to long range moves and emerging small world property
increased efficiency compared to a normal walk strategy.
These results confirm the recent findings in \cite{riascos-fracdyn,riascos-fracdiff}.

\section{Conclusion}

We analyzed time-discrete and time-continuous `fractional' random walks on undirected regular networks with special emphasis on cubic lattices in $n=1,2,3,..$ dimensions.
The random walk of our model is described by a master equation involving {\it fractional powers of Laplacian matrices $L^{\frac{\alpha}{2}}$}.
The fractional network Laplacian matrix represents the matrix analogue to the continuous fractional Laplacian (Riesz fractional derivative).
We demonstrated that the fractional random walk is admissible in the same range $0<\alpha\leq 2$ of power law index as L\'evy flights.

We derived analytical expressions for the fractional transition matrix, fractional fundamental matrix (Green's matrix)
and obtained representations for the return probabilities, the mean relaxation time and closely related the Kemeny constant.
We show for the cubic lattice explicitly the emergence of L\'evy flights for sufficiently large lattices.
For the return probabilities we obtain characteristic  $t^{-\frac{n}{\alpha}}$ power law decay for fractional walks in $n$-dimensional cubic lattices.
This scaling is the same as one obtains for fractional diffusion (L\'evy flights) in the continuous $n$-dimensional space \cite{michel-et-al-13}.
The fractional random walk model can be conceived as a model for anomalous diffusion on regular networks such as cubic lattices.
It can be concluded that, due to long range moves, the efficiency to explore the lattice is increased when instead of a normal
random walk a fractional random walk is chosen.

The fractional random walk approach derived in this paper can be generalized for an analysis of anomalous transport phenomena
on more complex undirected networks. The subject of fractional random walks on regular undirected networks deserves further investigation. For instance
with the present framework further characteristic quantities such as the MPFT (mean first passage time) for regular undirected networks can be derived.

\section{Appendix}
\label{sign}
In this appendix we demonstrate that $0<\alpha\leq 2$ is an admissible interval for fractional random walks. To this end
let us consider the transition matrix
\begin{equation}
\label{good}
 W(\delta t) =1-\delta t L
\end{equation}
which requires non-negative matrix elements $0\leq W_{pq} \leq 1 $ together with the normalization condition \\ $\sum_{p=0}^{N-1}W_{pq}(\delta t)=1$.
This property is fulfilled
if $L$ is a `good' Laplacian matrix where  \\ \\ {\bf (i)} its non-diagonal elements must be non-positive
$L_{pq}= -A_{pq} \leq 0$ ($p\neq q$) and  as a consequence of translational symmetry \\  \\ {\bf (ii)} its diagonal elements $L_{pp}=K=\sum_{q=0}^{N-1}A_{pq} >0$ must be
be positive indicating the degree of the network.

Now the goal is to determine for which $\alpha$-range the fractional Laplacian matrix $L^{\frac{\alpha}{2}}$ has also the required
good properties {\bf (i)} and {\bf (ii)} when the Laplacian matrix $L$ has those good properties.
\\ \\
Let us first consider {\bf (i)}: (a) Assume $W(\delta t)$ is such a good transition matrix, then also all its powers $W^m(\delta t)=W(m\delta t)$ and hence also
the exponential  $W(t)=e^{-Lt}= \lim_{m\rightarrow\infty} ({\hat 1} -\frac{t}{m}L)^m$ is a good transition matrix, especially with
 $W_{pq}(\delta t)=-L_{pq}\delta t= A_{pq}\delta t \geq 0$ ($p\neq q$).
It follows that the elements of transition matrix $[e^{-Lt}]_{pq}$ are all positive and normalization remains conserved in time\footnote{Since
$\frac{d}{dt}\sum_{q=0}^{N-1}W_{pq}(t)= -\sum_{s=0}^{N-1} \sum_{p=0}^{N-1}L_{ps}W_{sq} = 0$ only if $\sum_{p=0}^{N-1}L_{ps}=0$.}.
\\ \\
(b) Consider now for $\lambda \geq 0$ the integral

\begin{equation}
 \label{intref}
  \int_0^{\infty} t^{-\frac{\alpha}{2}-1} (e^{-\lambda t}-1) {\rm d}t = \lambda^{\frac{\alpha}{2}} C_{\alpha} ,\hspace{2cm} 0<\alpha < 2
\end{equation}
which exists for $0<\alpha <2$
with a {\bf negative} constant $C_{\alpha} <0$, occuring since the integrand for $\lambda >0$ is identically negative as $e^{-\lambda t}-1< 0$.
It can be seen by simple partial integration
that $C_{\alpha}=-\frac{2}{\alpha} \Gamma(1-\frac{\alpha}{2}) = \Gamma(-\frac{\alpha}{2}) < 0$ ($0<\alpha<2$).
\\ \\
(c) The fractional Laplacian matrix can then be represented by

\begin{equation}
 \label{fraclare}
 C_{\alpha} L^{\frac{\alpha}{2}} = \int_0^{\infty} t^{-\frac{\alpha}{2}-1} (e^{-Lt}-{\hat 1}) {\rm d}t
\end{equation}
which can be verified by employing the spectral representation of $L$ and (\ref{intref}).
The non-diagonal elements of the matrix (\ref{fraclare}) have the same sign as the non-vanishing elements of
$e^{-Lt}=W(t)$, namely $+1$, and remaining positive by the integration. We can conclude that the sign of (non-vanishing) non-diagonal elements of the fractional Laplacian matrix
$L^{\frac{\alpha}{2}}$ is determined by the (negative) sign of the constant $C_{\alpha}$, namely

\begin{equation}
\label{conclu}
sign([L^{\frac{\alpha}{2}}]_{pq}) = sign(C_{\alpha}) sign([e^{-Lt}]_{pq}) =(-1)\times 1 = -1  , \hspace{0.5cm} p\neq q , \hspace{0.5cm}  0<\alpha < 2
\end{equation}

Let us now analyze property {\bf (ii)} for the fractional Laplacian matrix. The positiveness of the in our case idential diagonal elements $L^{\frac{\alpha}{2}}_{pp}$ can be directly seen
from $L^{\frac{\alpha}{2}}_{pp}= \frac{1}{N}trace(L^{\frac{\alpha}{2}}) = \frac{1}{N} \sum_{\ell=0}^{N-1}\lambda_{\ell}^{\frac{\alpha}{2}} >0$. This further can be
confirmed by employing
(\ref{fraclare}), namely
\begin{equation}
 \label{diagfrac}
 L^{\frac{\alpha}{2}}_{pp}= \frac{1}{C_{\alpha}}\int_0^{\infty} t^{-\frac{\alpha}{2}-1}\left( \frac{1}{N} (\sum_{\ell=0}^{N-1}(e^{-\lambda_{\ell}t}-1)\right){\rm d}t ,
 \hspace{0.5cm} 0< \alpha <2
\end{equation}
and it follows again because of $ e^{-\lambda_{\ell}t}-1 <0 $ and $C_{\alpha} <0$ that $L^{\frac{\alpha}{2}}_{pp} >0$ in the interval of existence $0< \alpha <2$ of (\ref{diagfrac}).

The positive in general non-integer number $L^{\frac{\alpha}{2}}_{pp}= k^{(\alpha)}$ of the identical diagonal elements of the fractional Laplacian matrix
can be conceived as
{\it the fractional degree} in this interval. This interpretation was already suggested earlier \cite{riascos-fracdyn,riascos-fracdiff}.

In conclusion it follows that the non-diagonal matrix elements of the fractional Laplacian matrix holds equally as for $L$ the good property
$[L^{\frac{\alpha}{2}}]_{pq} <0 $ for $p\neq q$ and $L_{pp} >0$ in the interval $0<\alpha \leq 2$ where $\alpha=2$ recovers the good properties of $L$ .
The fractional Laplacian thus can be represented $0<\alpha\leq 2$

\begin{equation}
 \label{fractdlapgood}
 L^{\frac{\alpha}{2}}= k^{(\alpha)}{\hat 1}- A^{(\alpha)} , \hspace{2cm} 0<\alpha\leq 2
\end{equation}
 as suggested in \cite{riascos-fracdyn,riascos-fracdiff} generalizing normal random walks to fractional random walks.
 The fractional degree $k^{(\alpha)}= [L^{\frac{\alpha}{2}}]_{pp}$ and fractional analogue of adjacency
 matrix $A^{(\alpha)}_{pq} \geq 0$ was introduced there with $k^{(\alpha)} = \sum_{q \neq p} A^{(\alpha)}_{pq}$, having
the same good properties for the fractional case as for the normal walk case where $\alpha=2$. In the context of discrete fractional random walk
problem (\ref{modi}) the non-negative quantities $\frac{A^{(\alpha)}_{pq}}{k^{(\alpha)}}$ have the interpretations of transition probabilities.
This interpretation holds when $0\leq \frac{A^{(\alpha)}_{pq}}{k^{(\alpha)}}\leq 1$ which we showed to be fulfilled in the interval $ 0<\alpha\leq 2$.
For $\alpha >2$ it has been shown earlier for the
infinite cyclic chain, that alternating signs occur with oscillatoric behavior of the non-diagonal elements as a function of $\alpha$ \cite{michelJphysA}.

It follows that
the fractional Laplacian matrix approach yields
a good fractional transition matrix $W^{(\alpha)}(t)= e^{-L^{\frac{\alpha}{2}}t}$ only within the range $0 < \alpha \leq 2$. We emphasize that this is the same
$\alpha$-interval of existence of L\'evy flights and $\alpha$-stable L\'evy distributions.

\section*{Acknowledgements}
{\it  We are indepted to G.A. Maugin pour fruitful discussions.}


\begin{thebibliography}{1}
\bibitem{remark2}
As we deal with regular networks $K_p=K$, the re-normalization of the Laplacian matrix to $L_{pq}' =\frac{1}{K_p}L_{pq}$ which is often perfomed in the literature
is not needed here as does not provide any additional information and can be achieved by simple rescaling of the time step $\delta t \rightarrow K\delta t$.
\end{thebibliography}
\end{document}